\documentclass[oneside,12pt]{article}
\usepackage{epsfig}
\usepackage{latexsym}

\date{\ }

\newtheorem{definition}{Definition}
\newtheorem{theorem}{Theorem}
\newtheorem{proposition}{Proposition}

\newtheorem{lemma}{Lemma}
\newtheorem{example}{Example}

\newcommand{\fa}{$\cal F$}
\newcommand{\tr}{$\cal T$}
\newcommand{\un}{$\cal U$}
\newcommand{\inc}{$\cal O$}
\newcommand{\fas}{$\cal F$ }
\newcommand{\trs}{$\cal T$ }
\newcommand{\uns}{$\cal U$ }
\newcommand{\incs}{$\cal O$ }

\newcommand{\preuve}[1]{{\sc \small Proof. } #1 \hfill $\Box$ \\}

\title{ Hypotheses Founded Semantics of Logic Programs
for Information Integration  in 
Multi-Valued Logics\footnote{ A preliminary version of this paper
appeared in the form of an extended abstract in the proceedings of the first
                 International Conference on Theoretical Computer Science
		 (IFIP TCS 2000) - Exploring New Frontiers of
		 Theoretical Informatics.
                 }} 

\author{Yann Loyer\thanks{Corresponding author. Istituto di 
Elaborazione della Informazione, Consiglio Nazionale delle Ricerche, 
Area della Ricerca CNR di Pisa, Via Moruzzi,1 I-56124 Pisa, Phone (39) 050 315 
2901, Fax (39) 050 315 2810, {\tt loyer@lri.fr}} \and 
Nicolas Spyratos\thanks{Laboratoire de Recherche 
en Informatique, UMR
8623, Universit\'e de Paris Sud, 
         Bat. 490, 
        91405 Orsay, (33) 1 69 15 66 24, {\tt spyratos@lri.fr}. Part of this research
work was conducted while this author was visiting with the Meme Media
Laboratory, University of Hokkaido, Japan.}  
\and 
Daniel Stamate\thanks{Department of Computer Science, Birkbeck,
University of London, Malet Street, London WC1E 7HX, UK.
Phone +44-20 7631 6796, Fax +44-20 7631 6727, {\tt
d.stamate@dcs.bbk.ac.uk}.
Partially supported by a Marie Curie Fellowship, EC Contract
No: HPMF-CT-1999-0393}
}

\begin{document}

\maketitle
\begin{abstract}

We address  the problem of integrating information coming from 
different 
sources. The information consists of facts that 
a central server  collects  and tries to combine using (a) a set of 
logical 
rules, i.e. a logic program, and (b) a hypothesis representing the
server's own estimates. In such a setting  incomplete information 
from 
a source  or contradictory information from different sources 
necessitate the use of  many-valued logics in which programs can be 
evaluated and hypotheses can be tested. To carry out such activities 
we propose a formal framework based on bilattices such as  Belnap's 
four-valued logics. In
this framework  we work with the class of programs defined by Fitting
and we develop a theory for information integration. 
 We also establish an intuitively appealing connection between our
hypothesis testing mechanism on the one hand, and  the well-founded
semantics  
 and  Kripke-Kleene semantics of Datalog programs with negation, on
the other hand. 

\end{abstract}

\section{Introduction}

In several information oriented activities  there is a need for 
combining (or ``integrating'') information coming from different 
sources.

A typical example of such information-oriented activity is building a 
data warehouse, i.e. a special kind of  very large database for 
decision-making support in big enterprises \cite{acm}. The 
information stored in a data warehouse is obtained from queries to 
operational databases internal to the enterprise, and from remote 
information sources external to the enterprise  accessed through the 
Internet. The answers to all such queries are then combined (by the 
so-called ``integrator'') to derive the information to be stored in 
the data warehouse.

The basic pattern of the data warehouse paradigm, i.e. collection of 
information then integration, is encountered in many different 
situations. What changes usually from one situation to another is the 
type (and volume) of the collected information  and the means used 
for the integration.\\

 In this paper  we address a specific problem of  information 
integration, namely, the information consists of facts that a central 
server collects from a number of autonomous sources and then tries  
to  combine them using:
\begin{itemize}
\item a set of logical rules, i.e. a logic program, and
\item a hypothesis, representing the server's own estimates.
\end{itemize}

 In such a setting  incomplete information from a source or 
contradictory information coming from different sources necessitate 
the use of many-valued logics, in which programs can be evaluated and 
hypotheses can be tested. Let us see a simple example.

\begin{example} 
{\normalfont
Consider a legal case where a judge (the ``central server'') has to 
decide whether to charge a person named John accused of murder. To do 
so, the judge first collects facts from two different sources:  the 
public  prosecutor and the person's lawyer. The judge then combines 
the collected facts using a set of rules in order to reach a 
decision. For the sake of our example  let us suppose that the judge 
has collected a set of facts $F$ that he combines using a set of 
rules $R$ as follows:

$$ 
 F~~~
\left[
\begin{tabular}{cc}
\mbox{witness}(John)& false\\
 \mbox{friends}(John, Ted)& true
\end{tabular}
\right]\footnote{That notation means that the only facts which are 
defined (or assigned to logical values different from $unknown$) are 
witness(John) which is false and friends(John, Ted) which is true.}
$$

$$
 R~~~
\left\{
\begin{array}{lll}
  \mbox{suspect(X)} &  \leftarrow  & \mbox{motive}(X) \vee 
\mbox{witness}(X)\\
  \mbox{innocent}(X) &  \leftarrow  & \exists Y (\mbox{alibi}(X,Y) 
\wedge \neg \mbox{friends}(X,Y))\\
  \mbox{friends}(X,Y) &  \leftarrow  & \mbox{friends}(Y,X) \vee 
(\mbox{friends}(X,Z) \wedge \mbox{friends}(Z,Y))\\
\mbox{charge}(X) &   \leftarrow  &   \mbox{suspect}(X)  \oplus  \neg 
\mbox{innocent}(X)
\end{array}
\right.
$$

The first fact of $F$ says that there is no witness, i.e. the fact 
$\mbox{witness}(John)$ is false. The second fact of $F$ says that Ted 
is a friend of John, i.e. the fact $\mbox{friends}(John, Ted)$ is 
true.

Turning now to the set of rules, the first rule of $R$ describes how  
the prosecutor works: in order to support the claim that a person $X$ 
is a suspect, the prosecutor tries to provide a motive and/or a 
witness.

The second rule of $R$ describes how the lawyer works: in order to 
support the claim that  $X$ is innocent, the lawyer tries to provide 
an alibi for $X$ by a person who is not a friend of $X$. This rule   
depends on the third  rule which defines the relation $friends$.

Finally, the fourth rule of $R$ is the ``decision making rule'' and 
describes how the judge works: in order to reach a decision as to 
whether to charge $X$, the judge examines the premises $suspect(X)$ 
and $\neg innocent(X)$. As explained earlier, the values of these 
premises come from two different sources: the prosecutor and the 
lawyer. Each of these premises can have the value true or false. 
However, it is also possible that the value of a premiss is 
undefined. For example, if a motive is not known and a witness has 
not been found, then the value of suspect(X) will be undefined.

In view of these observations, the question is what value is 
appropriate to associate with charge(X).

 What we propose is to collect together the values of the premises 
suspect(X) and $\neg \mbox{innocent(X)}$, and to consider the 
resulting set of values as the value of charge(X). This is precisely 
what the notation
\vspace{-.1cm}
 $$\mbox{charge}(X)   \leftarrow \mbox{suspect}(X)  \oplus  \neg 
\mbox{innocent}(X)$$
means, where $\oplus$ denotes the ``collecting together'' operation.

It follows that there are four possible values for charge(X) : 
$\emptyset$, \{true\}, \{false\} and \{true, false\}. We shall call 
these values : $Underdefined$, $True$, $False$ and $Overdefined$, and 
we shall denote them by $\cal U$, $\cal T$, $\cal F$ and $\cal O$, 
respectively.

The value $Underdefined$ for a premiss means that the premiss is true 
or false but its actual value is currently unknown. For the purpose 
of this paper we shall assume that any premiss whose value is not 
known is associated with the value $Underdefined$.

We note that the value $Underdefined$ is related to the so-called 
``null values'' of attributes in database theory. In database theory, 
however, a distinction is made between two types of null values 
\cite{zan}: 
\begin{itemize}
\item the attribute value exists but is currently unknown;
\item the attribute value does not exist.
\end{itemize}

An example of the first type is the Department-value for an employee 
that has just been hired but has not yet been assigned to a specific 
department, and an example of the second type is the maiden name of a 
male employee. The value $Underdefined$ corresponds to the first type 
of null value.

Returning now to our example, the decision whether to charge John  
depends on the value that charge(John) will receive when collecting 
the values of the premises together. Looking at the facts of $F$ and 
the rules of $R$ (and using intuition)  we can see that suspect(John) 
and innocent(John) both receive the value $\cal U$  and so then does 
charge(John).

 This is clearly a case where the judge cannot decide whether to 
actually charge John! 

}

\end{example}

In the context of decision making, however, one has to reach a 
decision (based on the available facts and rules) even if some values 
are not defined. This can be accomplished by $assuming$ values for 
some or all underdefined premises. Such an assignment of values to 
underdefined premises is what we call a $hypothesis$.

Thus in our example, if the judge assumes the innocence of John, then 
charge(John) receives the value false and John is not charged. We 
note that this is precisely what happens in real life under similar 
circumstances, i.e. the defendant is $assumed$ innocent until proved 
guilty.

Clearly, when hypothesizing on underdefined premises we would like 
our hypothesis to be ``reasonable'' in some sense, with respect to 
the available information, i.e., with respect to the given facts and 
rules. Roughly speaking, we define a hypothesis $H$ to be 
``reasonable'' or $sound$ using the following test :\\

calling a fact $f$ $defined$ under $H$ if $H(f) \not = \cal U$,
 \begin{enumerate}
\item add $H$ to $F$ to produce a new set of facts $F' = F  \cup H$;
\item apply the rules of $R$ to $F'$ to produce a new assignment of 
values  $H'$;
\item if the facts defined under $H$ are assigned to the same values 
in $H'$ then $H$ is sound, otherwise $H$ is not sound.
\end{enumerate}

That is, if there is no fact of $H$ that has changed value as a 
result of rule application then $H$ is a sound hypothesis; otherwise 
$H$ is unsound.

In our example, for instance,  consider the following hypothesis:

$$
 H_1=
\left[
\begin{tabular}{cc}
\mbox{innocent}(John)&$\cal T$ \\
\mbox{charge}(John) &$\cal T$ 
\end{tabular}
\right]
$$

Applying the above test we find the following values for the facts of 
$H_1$ :

$$
 H'_1=
\left[
\begin{tabular}{cc}
\mbox{innocent}(John)&  $\cal T$\\
\mbox{charge}(John)&$\cal F$ 
\end{tabular}
\right]
$$

As we can see, the fact charge(John) has changed value, i.e. this 
fact had the value $\cal T$ in $H_1$ and now has the value $\cal F$ 
in $H'_1$. Therefore, $H_1$ is not a sound hypothesis.

Next, consider  the following hypothesis:

$$
 H_2=
\left[
\begin{tabular}{cc}
\mbox{innocent}(John)& $\cal T$\\
 \mbox{charge}(John)&$\cal F$ 
\end{tabular}
\right]
$$

Applying again our test we find :

$$
 H'_2=
\left[
\begin{tabular}{cc}
\mbox{innocent}(John)& $\cal T$\\
 \mbox{charge}(John)&$\cal F$ 
\end{tabular}
\right]
$$

That is, the values of the facts of $H_2$ remain unchanged in $H'_2$, 
thus $H_2$ is a sound hypothesis.

Intuitively, if our  hypothesis is sound this means that what we have 
assumed  is compatible with the given facts and rules.

From now on let us denote $\cal P$ the facts of $F$ together with the 
rules of $R$, i.e. ${\cal P} = \langle F,  R \rangle$, and let us 
call $\cal P$ a program.

In principle, we may assume or hypothesize values for every possible 
ground atom. However, given a program $\cal P$ and a hypothesis $H$, 
we cannot expect $H$ to be sound with respect to $\cal P$, in 
general. What we can expect is that some ``part'' of $H$ is sound 
with respect to $\cal P$.

More precisely,  given two hypotheses $H$ and $H'$, call $H$ a {\it 
part of} $H'$, denoted $H \leq H'$, if $H(f) \not = {\cal U}$ implies 
$ H(f) = H'(f)$, i.e., if $H$ agrees with $H'$ on every defined fact. 
It is then natural to ask, given program $\cal P$ and hypothesis $H$, 
what is the maximal part of $H$ that is sound with respect to $\cal 
P$. We call this maximal part the $support$ of $H$ by $\cal P$, and 
we denote it by $s_{\cal P}^H$. 
Intuitively, the support of $H$ indicates how much of $H$ can be 
assumed safely, i.e., remaining compatible with the facts and rules 
of $\cal P$.

We show that the support $s_{\cal P}^H$ can be used to define a 
hypothesis-based semantics of ${\cal P}=\langle F, R \rangle$, 
denoted by $sem^H_{\cal P}$. This is done by a fixpoint computation 
that uses an immediate consequence operator $T$ as follows:

\begin{itemize}
\item $F_0 = F$;
\item $F_{i+1} = T(F_i) \oplus s^H_{\langle F_i, R \rangle}$.
\end{itemize}
    
We also show that there is an interesting connection between  
hypothesis based semantics and the semantics of  Datalog programs 
with negation. More precisely, we show that if $\cal P$ is a Datalog 
program with negation  then:

\begin{itemize}
\item if $H$ is the everywhere false hypothesis  then $sem^H_{\cal 
P}$ coincides with the well-founded semantics of $\cal P$ 
\cite{gelder2}, and
\item if $H$ is  the everywhere underdefined hypothesis  then 
$sem^H_{\cal P}$  coincides with the Kripke-Kleene semantics of $\cal 
P$ \cite{fitting:kk}.
\end{itemize}

 As we shall see, these results allow us to extend the well-founded 
semantics and the Kripke-Kleene semantics of  Datalog program with 
negation to the broader class of Fitting programs \cite{fitting}.

Motivation for this work comes from the area of knowledge 
acquisition, where contradictions  may occur during the process of 
collecting knowledge  from different experts. Indeed, in   
multi-agent systems, different agents may give different answers to 
the same query. It is then important to be able to process the 
answers so as to extract the maximum of information on which the 
various agents agree, or to detect the items on which the agents give 
conflicting answers.

Motivation also comes from the area of  deductive databases. Updates 
leading to a certain degree of inconsistency should be allowed 
because inconsistency can lead to  useful information, especially 
within the framework of distributed databases. In particular,  Fuhr 
and R\"{o}lleke showed in \cite{fr97} that hypermedia information retrieval 
requires the handling of inconsistent information and non-uniform 
hypotheses.

The remaining of the paper is organized as follows. In Section 2  we 
recall very briefly some definitions and notations from  well-founded 
semantics,  Belnap's logic $\cal FOUR$, bilattices  and Fitting programs. We then 
proceed, in Section 3, to define sound hypotheses and their support 
by  a  Fitting program~$\cal P$; we also discuss computational issues 
and we present  algorithms for computing the support of a hypothesis 
by a program $\cal P$ and the hypothesis-founded semantics of $\cal P$. 
In Section 4 we show that the notion of support actually unifies the 
notions of well-founded semantics and Kripke-Kleene semantics and 
extends them from Datalog program with negation to the broader class 
of Fitting programs.
Section 5 contains concluding remarks and suggestions for further 
research.

\section{Preliminaries}

\subsection{Three-valued logics}

\subsubsection{Well founded semantics}
\label{wfs}

Well-founded semantics of logic programs were first proposed in 
\cite{gelder2}. The well founded semantics of a Program $\cal P$ is based on the 
closed  world assumption, i.e. every atom is supposed to be  $false$ by 
default. In the approach of \cite{gelder2}  an interpretation 
$I$  is a set of 
ground literals  that  does not contain  literals
of the form $A$ and $\neg A$. Now, if we consider an instantiated 
program $P$ defined as in \cite{gelder2}, its 
well-founded semantics is defined using the following two operators 
on partial interpretations $I$ :
\begin{itemize}
\item  the immediate consequence operator 
$T_P$, defined by $$T_P(I)=\{ head(r)~|~r\in P \wedge \forall B \in 
body(r), B\in I\}\mbox{, 
and}$$
\item the unfounded operator $U_{P}$, where $U_{P}(I)$ is defined to 
be the greatest unfounded 
set with respect to the partial interpretation 
$I$.  
\end{itemize}

We recall that a set of instantiated atoms $U$ is said to be 
unfounded with respect to $I$ 
if for all instantiated atoms $A\in U$ and for all rules $r\in {P}$ 
the following 
holds: $$head(r)=A \Rightarrow \exists B\in body(r)~( \neg B\in I 
\vee B \in U)$$

In \cite{BiF91} it is proven that $U_{P}(I) = {\cal HB}\setminus  
SPF_{P}(I)$, where ${\cal HB}$ is the Herbrand Base and $SPF_{P}(I)$ 
is the limit of the increasing sequence 
$[SPF^i (I)]_{i\geq 1}$ defined by: \vspace{.3cm}

\begin{tabular}{llll} 
$-~SPF^{1} _{P}(I)$&$=$&$\{ 
head(r)~|~r\in {P}$ &$\wedge  pos(body(r))=\emptyset$\\ 
&&&$\wedge \forall B\in body(r), 
\neg B \not \in~I\}$\\\\ 
$-~SPF^{i+1} _{P}(I)$&$=$&$\{ head(r)~|~r \in {P}$&$ \wedge  
pos(body(r)) \subseteq SPF^{i} _{P}(I)$\\
&&&$\wedge \forall B\in body(r),  \neg B \not \in  I\}, i>0.$
\end{tabular} \vspace{.3cm}

The atoms of $SPF_{P}(I)$ are called potentially founded atoms.

 The operator $W_{P}$, called the 
well-founded operator, is then defined by 
$W_{P}(I)=T_{P}(I)\cup \neg U_{P}(I)$ and is shown to be monotone 
with respect to set inclusion. The 
well-founded semantics of $P$ is defined to be the least fixpoint of 
$W_{P}$ 
\cite{gelder2}.

\subsubsection{Kripke-Kleene semantics}

The Kripke-Kleene semantics was introduced in \cite{fitting:kk}. In 
the approach of \cite{fitting:kk}, a valuation is a function from the 
Herbrand base to the set of logical values 
$\{true,~false,~unknown\}$. Now,  given  an instantiated program 
$\cal P$ defined as in \cite{fitting:kk}, its Kripke-Kleene semantics 
is defined using an operator  $\Phi_{\cal P}¥$  on valuations,   defined as 
follows :
given   a ground atom $A$,
\begin{itemize}
\item if there is a rule in  $\cal P$ with head  $A$, and the truth 
value of the body under $v$ is  $true$, then ${\Phi}_{\cal P} (v)(A) 
= true$;
\item  if there is a rule in  $\cal P$ with head $A$, and for every 
rule in $\cal P$ with head $A$ the truth value of the body under $v$ 
is false, then  ${\Phi}_{\cal P} (v)(A) = false$;
\item else  ${\Phi}_{\cal P} (v)(A) = unknown$.
\end{itemize}

The Kripke-Kleene semantics of a Program $\cal P$ is based on the 
open world assumption, i.e. every atom is supposed to be  $unknown$ by 
default, and is defined to be
is the iterated
fixpoint of $\Phi_{\cal P}¥$ obtained by beginning the iteration with 
the everywhere unknown valuation.

\subsection{Multi-valued logics}

\subsubsection{Belnap's four-valued logic}
In \cite{belnap:four}, Belnap defines  a logic called $\cal FOUR$ 
intended to deal with incomplete and inconsistent information. 
Belnap's logic uses four logical values that we shall denote by \fa, 
\tr, \uns and \incs, i.e. $\cal FOUR$ = \{\fa, \tr, \un, \inc\}. 
These values can be compared using two orderings, the knowledge 
ordering and the truth ordering.

In the knowledge ordering, denoted by ${\leq}_k$, the four values are 
ordered as follows: \uns ${\leq}_k$ \fa, \uns ${\leq}_k$ \tr, \fas 
${\leq}_k$ \inc, \trs ${\leq}_k$ \inc. Intuitively, according to this 
ordering, each value of $\cal FOUR$ is seen as a possible knowledge 
that one can have about the truth of a given statement. More 
precisely, this knowledge is expressed as a set of classical truth 
values that hold for that statement. Thus, \fas is seen as 
\{$false$\},  \trs is seen as \{$true$\}, \uns is seen as $\emptyset$ 
and \incs is seen as \{$false$,$true$\}. Following this viewpoint, 
the knowledge ordering is just the set inclusion ordering.

In the truth ordering, denoted by ${\leq}_t$, the four logical values 
are ordered as follows: \fas ${\leq}_t$ \un, \fas ${\leq}_t$ \inc, 
\uns ${\leq}_t$ \tr, \incs ${\leq}_t$ \tr.  Intuitively, according to 
this ordering, each value of $\cal FOUR$ is seen as the degree of 
truth  of a given statement. \uns and \incs are both less false than 
\fa, and less true than \tr, but \uns and \incs are not comparable.

The two orderings are represented in the double Hasse diagram of 
Figure~1.

\begin{figure}
\centering
        \hspace{-3cm} \setlength{\unitlength}{1500sp}%
{\begingroup\makeatletter\ifx\SetFigFont\undefined%
\gdef\SetFigFont#1#2#3#4#5{%
  \reset@font\fontsize{#1}{#2pt}%
  \fontfamily{#3}\fontseries{#4}\fontshape{#5}%
  \selectfont}%
\fi\endgroup%
\begin{picture}(3013,2270)(-200,-3000)
\thinlines
\put(1801,-1861){\circle*{150}}
\put(3001,-661){\circle*{150}}
\put(4201,-1861){\circle*{150}}
\put(3001,-3061){\circle*{150}}
\put(1501,-3661){\line( 1, 0){3000}}
\put(1201,-3361){\line( 0, 1){3000}}
\put(1201,-361){\makebox(1.6667,11.6667){\SetFigFont{5}{6}{\rmdefault}{\mddefault}{\updefault}.}}
\put(1801,-1861){\line( 1,-1){1200}}
\put(3001,-3061){\line( 1, 1){1200}}
\put(4201,-1861){\line(-1, 1){1200}}
\put(3001,-661){\line(-1,-1){1200}}
\put(10726,-6961){\makebox(1.6667,11.6667){\SetFigFont{5}{6}{\rmdefault}{\mddefault}{\updefault}.}}
\put(10726,-6961){\makebox(1.6667,11.6667){\SetFigFont{5}{6}{\rmdefault}{\mddefault}{\updefault}.}}
\put(1201,-361){\vector( 0, 1){ 75}}
\put(4426,-3661){\vector( 1, 0){150}}
\put(4351,-1936){\makebox(0,0)[lb]{\smash{\SetFigFont{6}{14.4}{\rmdefault}{\mddefault}{\updefault}{$\cal
T$}}}} \put(2701,-3361){\makebox(0,0)[lb]{\smash{\SetFigFont{6}{14.4}{\rmdefault}{\mddefault}{\updefault} 
$\cal U$}}}
\put(1351,-1936){\makebox(0,0)[lb]{\smash{\SetFigFont{6}{14.4}{\rmdefault}{\mddefault}{\updefault}$\cal
F$}}} \put(2776,-436){\makebox(0,0)[lb]{\smash{\SetFigFont{6}{14.4}{\rmdefault}{\mddefault}{\updefault}
$\cal O$}}}
\put(4201,-4036){\makebox(0,0)[lb]{\smash{\SetFigFont{6}{14.4}{\rmdefault}{\mddefault}{\updefault}$\leq_t$}}}
\put(601,-511){\makebox(0,0)[lb]{\smash{\SetFigFont{6}{14.4}{\rmdefault}{\mddefault}{\updefault}$\leq_k$}}}
\end{picture}}

        \vspace{1cm}
	\caption{The logic {\sl FOUR}
        \label{four}}
\end{figure}
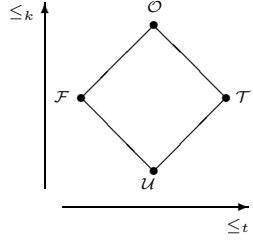

Both $\leq_t$ and $\leq_k$ give  $\cal FOUR$  a lattice structure. 
Meet and join under the truth ordering are denoted by $\wedge$ and 
$\vee$, and they are natural 
generalizations of the  usual notions of  conjunction and 
disjunction. In particular, \un $\wedge$\inc = \fas and \un 
$\vee$\inc = \tr. Under the knowledge ordering, meet and join are 
denoted by $\otimes$
and $\oplus$, and are called the $consensus$ and $gullibility$, 
respectively: $x
\otimes y$ represents the maximal information on which $x$ and  $y$  
agree, whereas $x \oplus
y$ adds the knowledge represented by  $x$ to that represented by  
$y$. In particular, \fa $\otimes$\tr = \uns and \fa $\oplus$\tr = 
\inc.

There is a natural  notion of negation in the truth ordering denoted 
by $\neg$, and we have: $\neg$ \tr = \fa, $\neg$ \fa = \tr, $\neg$ 
\un = \un, $\neg$ \inc = \inc. There is a similar  notion
for the knowledge ordering, called  $conflation$,
denoted by  -, and we have: - \un = \inc, - \inc = \un, - \fa = \fa, 
- \tr = \tr.

The operations $\vee, \wedge, \neg$ restricted to the values \trs and 
\fas are those of classical logic, and if we add to these operations 
and values the value \uns  then they are those of Kleene's strong 
three-valued logic.\\

\subsubsection{Bilattices}

In \cite{fitting:bilat,Mes97},  
bilattices are used as truth-value spaces for integration of
information coming from different sources.
The bilattice approach is a basic contribution to
many-valued logics. Bilattices and their derived sublogics
are useful in expressing uncertainty and inconsistency
in logic programming and databases 
\cite{arieliavron98,fitting:bilat,fitting,Gargov,MPS97,SS97}. 
 The simplest non-trivial 
bilattice is called {\sl FOUR}, and it is basically Belnap's
four-valued logic \cite{belnap:four}.

\begin{definition}
A bilattice is a triple $\langle {\cal B},\leq_{t},\leq_{k}\rangle$, 
where $\cal{B}$ 
is a nonempty set and $\leq_t$, $\leq_k$ are each a partial ordering 
giving 
$\cal{B}$ the structure of a lattice with a top and a bottom.
\end{definition}

In a bilattice  $\langle {\cal B},\leq_{t},\leq_{k}\rangle$,
meet and join under $\leq_t$ are denoted $\vee$ and $\wedge$, and
meet and join under $\leq_k$ are denoted $\oplus$ and $\otimes$. Top 
and
bottom under $\leq_t$ are denoted ${\cal T}$ and ${\cal F}$, and top
and
bottom under $\leq_k$ are denoted $\cal I$ and $\cal U$. 
If the bilattice is complete with respect to both orderings, 
infinitary meet and join under $\leq_t$ are denoted $\bigvee$ and 
$\bigwedge$, and infinitary 
meet and join under $\leq_k$ are denoted $\bigoplus$ and~$\bigotimes$.

\begin{definition}
A bilattice  $\langle {\cal B},\leq_{t},\leq_{k}\rangle$ is called  
distributive if all 12 distributive laws connecting $\vee$, $\wedge$, 
$\oplus$ and $\otimes$ hold. It is called infinitely distributive if 
it is a complete bilattice  in which all infinitary, as well as 
finitary, distributive laws  hold. 
\end{definition}

An example of a distributive law is $x\otimes(y\vee z)=(x\otimes y) 
\vee (x\otimes z)$. An example of an infinitary distributive law is 
$x\otimes \bigvee \{y_i | i \in S\} = \bigvee \{x \otimes y_i | i \in 
S\}$.

\begin{definition}
A  bilattice  $\langle {\cal B},\leq_{t},\leq_{k}\rangle$ satisfies 
the interlacing conditions if each of the operations $\vee$, 
$\wedge$, $\oplus$ and $\otimes$ is monotone with respect to both 
orderings. If the bilattice is complete, it satisfies the infinitary 
interlacing conditions if each of the infinitary meet and join is 
monotone with respect to both orderings.
\end{definition}

An example of an interlacing condition is:  $x_1 \leq_t y_1$ and $x_2 
\leq_t y_2$ implies $x_1 \otimes x_2 \leq_t y_1 \otimes y_2$. An 
example of an infinitary interlacing condition is: $x_i \leq_t y_i$ 
for all $i \in S$ implies $\bigoplus\{x_i | i \in S\} \leq_t 
\bigoplus\{y_i | i \in S\}$. A distributive bilattice satisfies the 
interlacing conditions.

$\cal FOUR$ is an infinitary distributive bilattice which satisfies 
the infinitary interlacing laws. A bilattice is said to be {\em 
nontrivial} if the bilattice 
{\sl FOUR} can be isomorphically embedded in it.\\

There are two principal ways for constructing  bilattices that were
introduced in  \cite{ginsberg:mv 
val}, and then developped in details in \cite{fitting:bilat}. 
The first one consists in considering two lattices $\langle L_1, \leq_1 \rangle$ and 
$\langle L_2, \leq_2 \rangle$. We can  see $L_1$ as the set of values 
used for representing the degree of belief (evidence, confidence, 
etc.) of an information and $L_2$ as the set of values used for 
representing the degree of doubt (counter-evidence, lack of 
confidence, etc.) of the information.  

We define the structure $L_{1} \odot L_{2}$ to be the structure 
$\langle L_1\times L_2,\leq_t,\leq_k\rangle$ where:
\begin{itemize}
\item $\langle x, y\rangle\leq_t\langle z,w\rangle$ iff
$x\leq z$ and $w\leq y$,\\
($\langle x, y\rangle\;\wedge\; \langle z,w\rangle$ =
$\langle min(x,z), max(y,w)\rangle$), and
\item $\langle x, y\rangle\leq_k\langle z,w\rangle$ iff
$x\leq z$ and $y\leq w$\\
($\langle x, y\rangle\;\otimes\; \langle z,w\rangle$ =
$\langle min(x,z), min(y,w)\rangle$).
\end{itemize}

$L_{1} \odot L_{2}$ 
is a bilattice satisfying the interlacing conditions; 
it is a complete bilattice satisfying the infinitary
interlacing conditions if $L_1$ and 
$L_2$ are complete;  it is infinitely distributive if $L_1$ 
and $L_2$ are complete and infinitely distributive. Moreover,
if $L=L_{1}=L_{2}$, then a negation can be defined by 
$\neg \langle x, y \rangle =\langle y, x \rangle$.

The following example illustrates possible uses of such a bilattice.

\begin{example}
Suppose that we have two information sources: two
veterinaries $v_1$ and $v_2$, and that we want to know the answer
to the query:
Is Marguerite  a crazy cow ?

 If
 $v_1$ asserts that the probability she is mad is 70\%, and $v_2$
asserts that the probability she is not  mad is 40\%, then this 
knowledge can be represented by assigning to the atom 
Mad(Marguerite) the logical value   $(0.7,0.4) \in [0;1] \times
[0;1]$. 

Such values could also be useful when each source can only answer by
true or false, but is associated to a specific degree of reliability.
The value $(0.7,0.4) $ could then  represent the fact that $v_1$
asserts that she is  mad  whereas $v_2$
asserts that she is not mad, but that we are more confident in 
diagnostics of $v_1$ than in those of  $v_2$. That  difference of 
reliability or confidence being represented by the assignation 
of different degrees of reliability to information sources, in our 
example,   $v_1$ would be  supposed
reliable for  $70\%$ and $v_2$ for $40\%$.
\end{example}    

The second way of constructing a bilattice consists in interpreting 
values as approximations of exact values. Suppose we have a lattice 
$\langle L, \leq_{L}\rangle$ of truth values. An approximation of a 
truth value can be seen as 
an interval $[a,b] = \{ x~|~a \leq_{L} x \leq_{L} b \}$ 
containing that value. We can provide to the set of intervals a  structure 
of bilattice $\langle {\cal O}(L), 
\leq_{t}, \leq_{k} \rangle$ such that, for $[a,b], [c,d] \in  {\cal 
O}(L)$:
\begin{itemize}
    \item $[a,b] \leq_{k} [c,d]$ if $a \leq_{L} c$ and $d \leq_{L} 
    b$, and
    \item $[a,b] \leq_{t} [c,d]$ if $a \leq_{L} c$ and $b \leq_{L} d$.
\end{itemize}

The intuition is that knowledge increases if the interval become 
shorter and truth increases if the interval contains greater values.\\

By abuse of notation we will sometimes talk about the bilattice
$\cal B$ when the orders are  irrelevant or understood from the 
context. 
From now on, we assume  that $\cal B$ is an infinitely 
distributive  bilattice that satisfy the infinitary 
interlacing conditions and has
a negation
 unless explicitly stated otherwise.

\subsubsection{Fitting  programs}

Conventional logic programming has the set \{\fa, \tr\} as its 
intended space of truth values  but since not every query may produce 
an answer  partial models are often allowed (i.e. \uns is added). If 
we want to deal with inconsistency  as well  then \incs must be 
added. Thus Fitting asserts that $\cal FOUR$ can be thought as the 
``home'' of ordinary logic programming and extends the notion of 
logic program, as follows:

\begin{definition}{\bf (Fitting program)}

\begin{itemize}
  \item  A formula is an expression built up from literals and 
elements of $\cal B$, using $ \wedge , \vee , \otimes , \oplus , 
\exists , \forall $. 
  \item A clause is of the form $P(x_1,...,x_n) \longleftarrow \phi 
(x_1,...,x_n)$, where the  atomic formula  $P(x_1,...,x_n)$ is the 
head, and  the formula $\phi (x_1,...,x_n)$ is the body. It is 
assumed that the free variables of the body are among   $x_1,...,x_n$.
  \item  A  program  is a finite set of clauses with no predicate 
letter appearing in the head of more than one clause (this apparent 
restriction causes no loss of generality \cite{fitting:bilat}).
\end{itemize}
\end{definition}

 We shall represent  a {\it Fitting program} as a pair $\langle F, R 
\rangle$ where $F$ is a function from the Herbrand base into $\cal 
B$  and $R$ a set of clauses. This is possible because every fact 
can be seen as a rule of the form $A \leftarrow v$, where $A$ is an 
atom and $v$ is a value in $\cal B$.
 
 A Datalog program with negation can be seen as a Fitting program   
whose underlying truth-value space is the subset $\{ \cal F, T, U \}$ 
of $\cal B$  and which does not involve $\otimes, \oplus, \forall, 
\cal U, \cal O, F$.

\section{Hypothesis Testing}

In the remaining of this paper,  in order to simplify the 
presentation, we assume that all Fitting programs are instantiated 
programs. Moreover, we use the term ``program'' to mean ``Fitting 
program'', unless explicitly stated otherwise.

 \subsection{Interpretations}

First, we introduce some terminology and notation that we shall use 
throughout the paper. Given a program $\cal P$, call {\it 
interpretation} of $\cal P$ any function $I$ over the Herbrand base 
$\cal {\cal HB}_P$ such that, for every atom $A$ of $\cal {\cal 
HB}_P$,  $I(A)$ is a value from $\cal B$.

 Two  interpretations $I$ and $J$ are {\it compatible} if, for every 
ground atom A, $(I(A) \not = {\cal U} ~\mbox{and}~ J(A) \not = {\cal 
U}) \Rightarrow I(A) = J(A)$.

An interpretation $I$ is a {\it part of} an interpretation $J$, 
denoted $I \leq J$, if $I(A) \not = {\cal U}$ implies $ I(A)=J(A)$, 
for every ground atom $A$. Clearly, the part-of relation just defined 
is a partial ordering on the set $\cal V(B)$ of all 
interpretations over $\cal B$. Given an interpretation $I$, we 
denote by $def(I)$ the set of all ground atoms $A$ such that 
$I(A)\not = \cal U$. Moreover, if $S$ is any set of ground atoms, we 
define the $restriction$ of $I$ to $S$, denoted by $I_{/S}$ as 
follows: for all $A \in {\cal HB}_{\cal P}$,
%\vspace{-.15cm}
$$
I_{/S}(A)=
\left\{
\begin{array}{l}
I(A)~~\mbox{if}~A \in S,\\
{\cal U}\mbox{, otherwise.}
\end{array}
\right.
$$

We can extend the two orderings of ${\cal B}$ (i.e. the truth 
ordering and the knowledge ordering) to the set  $\cal V(B)$ as 
follows:
Let $I_1$ and $I_2$ be in $ \cal V$($\cal B$), then
\begin{itemize}
\item $I_1 \leq_t I_2$ if and only if $I_1(A) \leq_t I_2(A)$ for all 
ground atoms $A$;
\item
$I_1 \leq_k I_2$  if and only if $I_1(A) \leq_k I_2(A)$ for all 
ground atoms $A$.
\end{itemize}

Under these two orderings  $ \cal V$($\cal B$) becomes a 
bilattice, and we have \\$(I~\wedge~J)(A)~=~I(A)~\wedge~J(A)$, and 
similarly  for the other operators. $ \cal V$($\cal B$) is  
distributive,  satisfies the  interlacing conditions and has a 
negation and a conflation. \\

The operations of $\cal B$ can be extended naturaly to $\cal 
V(B)$ in the following way: $I \oplus J(A)= I(A) \oplus J(A)$ and 
similarly for the other operations.

The actions of  interpretations can be extended from atoms  to 
formulas as follows:
\begin{itemize}
\item $I(X \wedge Y) = I(X) \wedge I(Y)$, and similarly for the other 
operators,
\item $I((\exists x)\phi (x))= \bigvee_{t=closedterm} I(\phi (t)) $, 
and
\item $I((\forall x)\phi (x))= \bigwedge_{t=closedterm} I(\phi (t)) $.
\end{itemize}

If $B$ is a closed formula then we say that $B$  evaluates to the 
logical value
$\alpha$, with respect to an interpretation  $I,$ denoted by
 $B\equiv\alpha$ w.r.t. $I$ or by $B \equiv_I \alpha$, if 
$J(B)=\alpha$ for any interpretation $J$ such that $I \leq J$ (i.e. 
if the value of $B$ is equal to $\alpha$ with respect to the defined 
atoms of $I$ whatever the values of underdefined atoms could be). 
There are formulas $B$ in which
underdefined atoms do not matter for the logical value that can be 
associated with $B$. 
For example let us take $B=A\vee C$ and let the interpretation
$I$ be defined by $I(A)={\cal U},~I(C)=\cal T$; then no matter how 
$A$ is interpreted  $B$ is evaluated to $\cal T$,  that is, 
$B\equiv_I \cal T$. 
 
Given an interpretation $I$,  let $I_{{\cal O}}$ be the interpretation 
defined by : if   $I(A) \not =~\cal U$ then  $I_{{\cal O}}(A)= I(A)$ 
else  $I_{{\cal O}}(A)={\cal O}$, for every atom $A$.
Using the interlacing conditions, we have the following lemma 
that provides a method of testing whether
$B \equiv_I \alpha$, based on the interpretation $I_{{\cal O}}$.

\begin{lemma}\label{l1}
Given a closed formula $B$, $B \equiv_I \alpha$ iff $I(B) =\alpha$ 
and $I_{{\cal O}}(B)=\alpha$.
\end{lemma}

\subsection{The Support of a Hypothesis}

Given a program $P = \langle F, R \rangle$, we consider two ways of 
inferring  information from $\cal P$. First by activating the rules 
of $R$  in order to derive new facts from those of $F$, through an 
immediate consequence operator $T$. Second, by a kind of default 
reasoning based on a given hypothesis.  

\begin{definition}[immediate consequence operator $T$]
 The immediate consequence operator $T$   takes 
as  input the facts of $F$ and returns an interpretation $T(F)$, 
defined  as follows: for all ground atoms~$A$, 
\begin{itemize}
    \item if there is a rule  $ A\leftarrow B \in R$, then $
T_R(F)(A)= \alpha$ if $ B 
\equiv_F \alpha$,
\item $T_R(F)(A)=
\cal U$, otherwise.
\end{itemize}
\end{definition}

What we call a $hypothesis$ is actually just an interpretation
$H$. However, we use the term ``hypothesis'' to stress the fact that
the values assigned by $H$ to the atoms of the Herbrand base are
$assumed$ values - and $not$ values that have been computed using the
facts and rules of the program. As such, a hypothesis $H$ must be
tested against the ``sure'' knowledge provided by  $\cal P$.  The test
consists of ``adding''  $H$ to $F$, then activating the rules of $\cal
P$ (using $T$) to derive an interpretation $H'$. If  $H\leq H'$, then
the hypothesis $H$ is a sound one, i.e.  the   values defined by  $H$
are not in contradiction with those defined by $\cal P$.  Hence the
following definition: 

\begin{definition}[Sound Hypothesis]
Let ${\cal P}=\langle F, R \rangle$ be  a  program and $H$ a
hypothesis.  $H$ is sound w.r.t. $\cal P$  if 
\begin{itemize}
\item $F$ and $H$ are compatible, and
\item  $ H_{/Head({\cal P})} \leq T(F \oplus H)$, where $Head(\cal P)=
\{A~|~\exists A \leftarrow B \in \cal P \}$. 
\end{itemize}
\end{definition}%\vspace{-0.15cm}

We use the restriction of $H$ to $Head({\cal P})$  before making the
comparison with $T(F \oplus H)$ because all atoms which are not head
of any rule of $\cal P$ will be assigned to the value $Underdefined$
by $T(F \oplus H)$. Then $H$ and $T(F \oplus H)$ are compatible on
these atoms.

The following example illustrates the definition of sound hypothesis
with respect to a logic program.

\begin{example}
\label{ex1}
We consider  the program $\cal P =\langle F,R \rangle$
defined  by :

\begin{center}
$
\mbox{ \it F }=
\left[
\begin{array}{cc}
\mbox{witness(Jean)}&
{\cal T}
\end{array}
\right]$\footnote{That notation means that the only atom which is 
assigned to a logical value different from  $\cal U$ in $F$ is witness(Jean)
and that its value is $\cal T$.}
\end{center}
\vspace{-.6cm}

%Cette notation signifie que le seul atome ayant une valeur
%diff\'erente de $\cal U$ dans $F$ est temoin(Jean) et que sa valeur
%est $\cal T$.

$$
 R~~~
\left\{
\begin{array}{lll}
  \mbox{suspect(X)} &  \leftarrow  & \mbox{motive}(X) \vee 
\mbox{witness}(X)\\
  \mbox{innocent}(X) &  \leftarrow  & \exists Y (\mbox{alibi}(X,Y) 
\wedge \neg \mbox{friends}(X,Y))\\
  \mbox{friends}(X,Y) &  \leftarrow  & \mbox{friends}(Y,X) \vee 
(\mbox{friends}(X,Z) \wedge \mbox{friends}(Z,Y))\\
\mbox{charge}(X) &   \leftarrow  &   \mbox{suspect}(X)  \oplus  \neg 
\mbox{innocent}(X)
\end{array}
\right.
$$
\vspace{.2cm}

 Let $H$ be the following  hypothesis  :

$$
\mbox{ \it H}=
\left[
\begin{array}{cc}
\mbox{witness(Jean)}&{\cal F}\\
\mbox{motive(Jean)}&{\cal F}\\
\mbox{suspect(Jean)}&{\cal F}\\
\mbox{innocent(Jean)} &{\cal T}

\end{array}
\right]
$$
\vspace{.2cm}

We can easily note that  H  is not sound with respect to  ${\cal
P}$. The atom 
witness(Jean) is defined in   $H$ and in $F$, but with
  different values, so $H$ and $F$ are not  compatible.

The maximal part of $H$ that is compatible with $F$ is
$$
\mbox{ \it H'}=
\left[
\begin{array}{cc}
\mbox{motive(Jean)}&{\cal F}\\
\mbox{suspect(Jean)}&{\cal F}\\
\mbox{innocent(Jean)} &{\cal T}

\end{array}
\right]
$$
\vspace{.2cm}

$F$ and $H'$ are compatible, so it is possible to collect the knowledge
defined by these two interpretation in a new one without creating 
conflicts or inconsistencies.

$$
F \oplus H' =
\left[
\begin{array}{cc}
\mbox{witness(Jean)}&
{\cal T}\\
\mbox{motive(Jean)}&{\cal F}\\
\mbox{suspect(Jean)}&{\cal F}\\
\mbox{innocent(Jean)} &{\cal T}
\end{array}
\right]
$$
\vspace{.2cm}

Then we activate the rules of $R$ on the interpretation $F \oplus
H'$  :

$$
T_R(F \oplus  H') =
\left[
\begin{array}{cc}
\mbox{witness(Jean)}&
{\cal T}\\
\mbox{motive(Jean)}&{\cal F}\\
\mbox{suspect(Jean)}&{\cal T}\\
\mbox{charge(Jean)}&
{\cal F}
\end{array}
\right]
$$
\vspace{.2cm}

We observe that 
$H'$ is not sound with respect to ${\cal P}$ because
$H'$ is not a part of $T_R(F
\oplus H)$ and is in contradiction with the derived knowledge.
\end{example}

Even if a hypothesis $H$ is not sound w.r.t.  $\cal P$, it may be that
some part of $H$ is sound w.r.t. $\cal P$. Of course, we are
interested to know what is the maximal part of $H$ that is  sound
w.r.t. $\cal P$. We shall call this maximal part the ``support'' of
$H$. To see that the maximal part of $H$ is unique (and thus that the
support is a well-defined concept), we give the following lemma: 
\begin{lemma}
If $H_1$ and $H_2$ are two sound parts of $H$ w.r.t. $\cal P$, then
$H_1 \oplus H_2$ is sound  w.r.t.~$\cal P$. 
\end{lemma}
\preuve{
$H_1$ and $H_2$ are both restrictions of $H$ so they are compatible.
Moreover, $H_1$ and $H_2$ are two sound parts of $H$ with respect to 
$\cal
P$, so $H_1$ and $H_2$ are both compatible with $F$. It follows that
$H_1 \oplus H_2$
is compatible with $F$. 

We also have $ {H_1}_{/\mbox{Heads}({\cal R})} \leq T(F \oplus
H_1)$, i.e., for all atom $A$ head of a rule   $A \leftarrow
B \in \cal P$,
    if $H_1 (A) \not = \cal U$ then
$H_1 (A) = T(F \oplus
H_1) (A)$. The same  property is verified by $H_2$.
If $H_1 \oplus H_2 (A)  \not = \cal U$ then we have :
\begin{itemize}
    \item either $H_1 (A)  \not =
\cal U$ and $ H_2 (A)   = \cal U$, 
\item either $ H_2 (A)  \not = \cal U$ 
and  $
H_1 (A)   = \cal U$, 
\item either $H_1 (A) = H_2 (A)  \not = \cal U$. 
\end{itemize}

In the first case, we have $T(F \oplus
H_1) (A) = H_1 (A)  = H_1 \oplus H_2 (A) $, 
i.e. \\ $B \equiv H_1 \oplus H_2 (A)
~p.r.~ F \oplus
H_1$, so $B \equiv H_1 \oplus H_2 (A)
~w.r.t.~ F \oplus
H_1 \oplus H_2$. We have the same result in the two other cases, so

$$ {(H_1 \oplus H_2) }_{/\mbox{Heads}({\cal R})} \leq 
T(F \oplus
H_1 \oplus H_2 ).$$

}

Thus the maximal sound part of $H$ is defined by $\bigoplus \{ H'~|~H'
\leq H~\mbox{and}~H'$ $\mbox{is sound w.r.t.}~\cal P\}$. 

\begin{definition}[Support]
Let $\cal P$ be  a  program and $H$ a  hypothesis. The support of $H$
w.r.t. $\cal P$, denoted  $s_{{\cal P}}^H$, is the maximal sound part
of $H$  w.r.t. $\cal P$ (where maximality is understood w.r.t. the
part-of ordering $\leq$). 

\end{definition}

\begin{example}
Let $\cal P$ be the program  and  $H$ the hypothesis defined in
the example \ref{ex1}, then the support of $H$ with respect to $\cal P$
is :
$$
 s_{{\cal P}}^H= 
\left[
\begin{array}{cc}
\mbox{motive(Jean)}&
{\cal F}
\end{array}
\right]
$$

\end{example}

We can remark that the  support of a hypothesis with respect to a program
${\cal P} = \langle R, F \rangle$ is compatible 
with the interpretation
obtained by activating the rules of $R$ on the facts of  $F$.

\begin{lemma}
\label{lemme-hypo-comp}
 Let ${\cal P} = \langle R, F \rangle$ be a logic  program and $H$
a  hypothesis. $T_R (F)$ and $s_{{\cal P}}^H$ are compatible. 
\end{lemma}
\preuve{
For all  atom A, if A is not the head of any rule of
$R$, then  $T_R (F)(A) \leq s_{{\cal P}}^H(A)$. If there is in  $R$ a
rule forme $A \leftarrow B$, then : 
\begin{itemize}
\item if $s_{{\cal P}}^H (A) =H(A)  \not = \cal U$, then $B \equiv 
\alpha ~w.r.t.~ F \oplus s_{{\cal
P}}^H$, and  if   $T_R(F)(A)  \not
= \cal U$, then $T_R(F)(A) = H(A)$ ;
\item  if   $T_R(F)(A) = \alpha \not
= \cal U$, then   $B \equiv \alpha ~
w.r.t.~ F$ and $B \equiv \alpha~ w.r.t.~ F \oplus s_{{\cal P}}^H$, 
and if $s_{{\cal P}}^H (A) \not = \cal U$, then $s_{{\cal
P}}^H (A)  = \alpha$.
\end{itemize}

}

We now give an algorithm for computing the support $s_{{\cal P}}^H$ 
of a hypothesis $H$ w.r.t. a program  $\cal P$.

Consider the following sequence $\langle PF_i \rangle, ~i\geq 0$:

\begin{itemize}
\item $PF_0 = \emptyset$;
\item$PF_{i} = \{ A ~|~ A \leftarrow B \in {\cal P}\,$ and $\, B
\not\equiv H(A) ~\mbox{\it w.r.t.} ~F \oplus H_{/({\cal  HB_P}\setminus 
IF(F,H))
\setminus PF_{i-1}}\}$ for all $i\geq 0$, \\ where $ IF(F,H)$ 
is teh set of facts that are imcompatible with  $H$, defined
by  $ IF(F,H) = \{ A~|~
(F(A) \not = {\cal U}) \wedge (H(A) \not = {\cal U}) \wedge  (F(A) 
\not =
H(A)) \}$.

\end{itemize}

%\begin{itemize}
%\item $PF_0 = \emptyset$;
%\item$PF_{i+1} = \{ A ~|~ A \leftarrow B \in {\cal P}\,$ and $\, B 
%\not\equiv H(A) ~w.r.t. ~F\oplus H_{/({\cal HB_P}\setminus 
%PF_i)\setminus def(F)}\}$ for all $i\geq 0$,

%\end{itemize}

%\begin{itemize}
%\item $PF_0(I) = \emptyset$;
%\item$PF_{i+1}(I) = \{ A ~|~ A \leftarrow B \in {\cal P}\,$ and $\, 
%B \not\equiv H(A) ~w.r.t. ~J \}$ for all $i\geq 0$,
 
% where J is the interpretation defined by: 
%$$
%J(A)=
%\left\{
%\begin{array}{l}
%I(A) ~\mbox{if}~ I(A) \not = {\cal U},\\
%H(A) ~\mbox{if}~ A\in (pos(B)\setminus PF_i)~\mbox{and}~I(A)={\cal 
%},\\
%{\cal U}\mbox{, otherwise.} 
%\end{array}
%\right.
%$$
%\end{itemize}

 The intuition here  is that 
we want to evaluate step by step the atoms that could potentially 
have a logical value different than their values in $H$. We have the 
following results:

\smallskip
\begin{proposition}
The sequence $\langle PF_i \rangle, ~i\geq 0$ is increasing with 
respect to set inclusion
and it has a limit reached in a finite number of steps. This limit 
is  denoted $PF$.
\end{proposition}
\preuve{
We show by  recurrence that for all $n$, $PF_{n-1} \subseteq
PF_n$.

$PF_0= \emptyset$ so the property is satisfied for  $n = 1$.\\

Suppose that $PF_{i-1} \subseteq
PF_i$. Thus we have $$H_{/({\cal HB_P} \setminus IF(F,H)) \setminus 
PF_i}
\leq H_{/({\cal HB_P} \setminus IF(F,H)) \setminus PF_{i-1}}.$$

  For all atom $A$, if there is in 
$\cal P$ a rule $A \leftarrow B$, then if $\, B
\not\equiv H(A)$  with respect to $F \oplus H_{/({\cal HB_P}
\setminus IF(F,H)) \setminus PF_{i-1}} $, then $\, B
\not\equiv H(A) $  with respect to $ F \oplus H_{/({\cal HB_P}
\setminus IF(F,H)) \setminus PF_{i}} $, and consequently, $PF_{i} = \{ A ~|~ A
\leftarrow B \in {\cal P}\,$ and $\, B
\not\equiv H(A) ~\mbox{\it with respect to} ~F \oplus H_{/({\cal HB_P}
\setminus IF(F,H)) \setminus PF_{i-1}}\} \subseteq \{ A ~|~ A
\leftarrow B \in {\cal P}\,$ and $\, B
\not\equiv H(A) ~\mbox{\it with respect to} ~F \oplus H_{/({\cal HB_P}
\setminus IF(F,H)) \setminus PF_{i}}\} =
PF_{i+1}$.

}

If an atom of the Herbrand base is not in   $PF$, then it means that, 
with respect to $\cal P$, 
there is no way of inferring  for that atom  a logical value different
than its value in 
$H$ w.
\begin{theorem}
Let $\cal P$ a logic program and $H$ a hypothesis, we have
$$s_{{\cal P}}^H = H_{/({\cal
HB_P}\setminus IF(F,H)) \setminus PF}$$  
\end{theorem}
\preuve{
We note $X = H_{/({\cal  HB_P}\setminus IF(F,H)) \setminus PF}$.
Firstly, we show that $X$ is a sound part of
$H$ with respect to $\cal P$. By definition, $X$ and $F$ are
compatible. Let $A$ be an atom such that there exists in  $R$ a rule
$A \leftarrow B$ and such that $X(A) =H(A) \not = \cal U$.
Then $A \not \in PF$ and $ B \equiv H(A) ~p.r.~ F \oplus X$, so $X$
is sound.

Secondly, we prove that $X$ is the maximal sound part of
$H$ with respect to  $\cal P$.
Let $Y$  be a sound part of $H$. We show by  recurrence 
that for all atom $A$, if $Y(A) = H(A) \not = \cal U$, then $A \not \in
PF_n$.

If $A$ is not the head of any rule in $R$, then $A 
\not \in
PF_i$, for all $i$. 
$PF_0 = \emptyset$ so $A \not \in
PF_0$.
Suppose the  property satisfied for $n = i-1$. We have, 
for all atom $A$, if $Y(A) = H(A) \not = \cal U$, then $A \not \in
PF_{i-1}$, and $Y \leq H_{/({\cal  HB_P}\setminus IF(F,H))
\setminus PF_{i-1}}$. 

If there is  a rule $A \leftarrow B$ in $R$ and $Y(A) = H(A) \not = \cal U$,
then\\  $B \equiv H(A) ~w.r.t.~ F \oplus  Y$ because $Y$ is sound, 
and it follows that\\  
 $B \equiv H(A) ~w.r.t.~ F \oplus H_{/({\cal  HB_P}\setminus IF(F,H))
\setminus PF_{i-1}}$. We can conclude that $A \not \in PF_i$.

We have shown that for all atom $A$, if $Y(A) = H(A) \not = \cal U$, 
then
$A \not \in
PF$ and consequently, $X(A) = H(A)$.

For all sound part  $Y$ of $H$, we have  $Y
\leq X$.

}

\section{Hypothesis Founded Semantics}

As we explained earlier, given a program $P = \langle F, R \rangle$, 
we derive information in two ways: by activating the rules (i.e.
by applying the immediate consequence operator T) and by making
a hypothesis $H$ and computing its support $s_{{\cal P}}^H$ w.r.t. 
$\cal P$. In the whole, the information that we derive comes from 
$T(F) \oplus s_{{\cal P}}^H$.

Now, roughly speaking, the semantics that we would like to associate 
with a program $\cal P$ is the maximum of information that we can 
derive from $\cal P$ under a sound hypothesis $s_{{\cal P}}^H$ but 
{\it without} any other information. To implement this idea we 
proceed as follows:
\begin{enumerate}
\item As we don't want any extra information (other than $\cal P$ 
 and $s_{{\cal P}}^H$), we use the everywhere undefined 
interpretation, call it $I_{\cal U}$.
\item In order to actually derive the maximum of information from 
$\cal P$ and $I_{\cal U}$, we collect together the knowledge 
infered by activating the rules of $R$, i.e. by applying the operator 
$T_{R}$, and as much of assumed knowledge as possible, i.e. 
the support of $H$ w.r.t. $\cal P$. 
\end{enumerate}

\begin{proposition} \label{p1}
The sequence $\langle F_n \rangle,~n\geq 0$ defined by :
\begin{itemize}
    \item $F_0 = F$  
, and
\item $F_{n+1}=T_R(F_n) \oplus s_{{\langle F_n, R \rangle}}^H$,
\end{itemize}
is 
increasing with
respect to $\leq$ and  has a limit denoted 
by $sem^H_{\cal P}$.
\end{proposition}
\preuve{ It is straighforward that  $T_R$ is monotonic with respect to
$\leq$, so for all n, $T_R(F_{n}) \leq T_R(F_{n+1})$.

We prove by recurrence that for all  n,  $s_{{\langle F_n, R
\rangle}}^H \leq s_{{\langle F_{n+1}, R 
\rangle}}^H$. 

For $n = 0$, if  $s_{{\langle F, R
\rangle}}^H (A) =\alpha \not = \cal U$ then :
\begin{itemize}
    \item if $A$ is not the head of 
any rule in $R$, then $s_{{\langle F_1, R
\rangle}}^H (A) =\alpha$ ;
\item if there is a rule $A
\leftarrow B$ in $R$, then $B \equiv \alpha ~w.r.t.~ s_{{\langle F, R
\rangle}}^H $, \\ so $B \equiv \alpha ~w.r.t.~ s_{{\langle F, R
\rangle}}^H \oplus T_R(F)$ and $B \equiv \alpha ~w.r.t.~ F_1$. \\ It 
follows that
$B \equiv \alpha ~w.r.t.~ F_1 \oplus s_{{\langle F_1, R
\rangle}}^H $. Consequently, $s_{{\langle F_1, R
\rangle}}^H = \alpha$.
\end{itemize}
The property is true for $n = 0$.\\\\
Now, suppose that $s_{{\langle F_{n-1}, R
\rangle}}^H \leq s_{{\langle F_{n}, R 
\rangle}}^H$. If $s_{{\langle F_{n}, R 
\rangle}}^H (A) = \alpha \not = \cal U$, then
\begin{itemize}
    \item  if $A$ is not the head of 
any rule in $R$, then $s_{{\langle F_{n}, R
\rangle}}^H (A) =\alpha$ ;

    \item  if there is a rule $A
\leftarrow B$ in $R$, then $B \equiv \alpha ~w.r.t.~ F_{n-1} \oplus 
s_{{\langle F_{n-1}, R
\rangle}}^H $.
\end{itemize}
 For all n, $F_{n} \oplus s_{{\langle F_{n}, R
\rangle}}^H \leq T_R(F_{n}) \oplus s_{{\langle F_{n}, R
\rangle}}^H$. Indeed, if $F_{n} \oplus s_{{\langle F_{n}, R
\rangle}}^H (A) = \alpha \not = \cal U$, then : 
\begin{itemize}
    \item  $T_R(F_{n-1}) 
(A)
= \alpha$, and it follows that $T_R(F_{n}) (A)
= \alpha$,\\ because $T_R(F_{n-1}) \leq T_R(F_{n}) $ ; or

    \item  $s_{{\langle 
F_{n-1}, R
\rangle}}^H (A) = \alpha$, and it follows that $s_{{\langle F_{n}, R
\rangle}}^H (A) = \alpha$,\\ because $s_{{\langle F_{n-1}, R
\rangle}}^H \leq s_{{\langle F_{n}, R
\rangle}}^H $ ; or

    \item  $s_{{\langle F_{n}, R
\rangle}}^H (A) = \alpha$.
\end{itemize}
 Consequently, $B \equiv \alpha ~w.r.t.~ T_R(F_{n-1}) \oplus
s_{{\langle F_{n-1}, R 
\rangle}}^H $, i.e. $B \equiv \alpha ~w.r.t.~F_{n}$, thus $B \equiv
\alpha ~w.r.t.~ F_{n} \oplus s_{{\langle F_{n}, R 
\rangle}}^H $. Finally $s_{{\langle F_{n}, R 
\rangle}}^H (A) = \alpha$.\\

So we have for all n, $T_R(F_{n}) \leq T_R(F_{n+1})$ and
$s_{{\langle F_n, R
\rangle}}^H \leq s_{{\langle F_{n+1}, R 
\rangle}}^H$, following the  lemma \ref{lemme-hypo-comp}, we can conclude that 
for all n, $F_{n} \leq F_{n+1}$.

}

\begin{proposition}
The interpretation $sem_{\cal P}^H$ is a model of $\cal P$.
\end{proposition}
\preuve{ $sem_{\cal P}^H = T_R(sem_{\cal P}^H) \oplus s_{{\langle
sem_{\cal P}^H, R \rangle}}^H$. But $sem_{\cal P}^H$ is a sound 
hypothesis, so, by definition, we have $${s_{{\langle
sem_{\cal P}^H, R \rangle}}^H}_{/\mbox{\small Heads(R)}} \leq 
T_R(sem_{\cal P}^H  \oplus s_{{\langle
sem_{\cal P}^H, R \rangle}}^H).$$
 It follows that
$${s_{{\langle 
sem_{\cal P}^H, R \rangle}}^H}_{/\mbox{\small T\^etes(R)}} \leq 
T_R(T_R(sem_{\cal
P}^H)  \oplus s_{{\langle 
sem_{\cal P}^H, R \rangle}}^H),$$ i.e. $${s_{{\langle 
sem_{\cal P}^H, R \rangle}}^H}_{/\mbox{\small T\^etes(R)}} \leq 
T_R(sem_{\cal
P}^H).$$

Thus, for all rule $A \leftarrow B \in R, sem_{\cal P}^H (A) =
T_R(sem_{\cal P}^H) (A) = \alpha$ if $B \equiv \alpha$ with respect to
$sem_{\cal P}^H$,  so  $sem_{\cal P}^H $ is  a model of $\cal
P$.

}

This justifies the following definition of semantics for $\cal P$.

\begin{definition}[Hypothesis founded semantics of $\cal P$]
The interpretation $sem_{\cal P}^H$
is defined to be the  semantics of $\cal P$ w.r.t. $H$ or
the $H$-founded semantics of $\cal P$. 
\end{definition}

Following this definition, any given program $\cal P$ can be 
associated with  different semantics, one for each possible 
hypothesis $H$. Theorem 2 below asserts that this approach extends 
the usual semantics of Datalog programs with negation to a broader 
class of programs, namely the Fitting programs.

Two remarks are in order here  before stating Theorem 2. First, if we 
restrict our attention to three values only, i.e. $\cal F$, $\cal T$ 
and $\cal U$, then our definition of interpretation is equivalent to 
the one used by Van Gelder et als \cite{gelder2}, in the following 
sense: given an interpretation $I$ following our definition, the set 
$\{A ~|~ I(A)={\cal T}\}\cup\{\neg A ~|~ I(A)={\cal F}\}$ is a 
partial interpretation following \cite{gelder2}; conversely, given a 
partial interpretation $J$ following \cite{gelder2}, the  function 
$I$ defined by: $I(A)=\cal T$ if $A \in J$, $I(A)=\cal F$ if $\neg A 
\in J$, and $I(A)=\cal U$ otherwise, is an interpretation in our 
sense. 

Second, if we restrict our attention to Datalog programs with 
negation (recall that the class of Fitting programs strictly contains 
the Datalog programs with negation)  then the concept of sound 
interpretation for the everywhere false hypothesis reduces to that of 
unfounded set of Van Gelder et als \cite{gelder2}. 
 The difference is that 
the definition in \cite{gelder2} has rather a syntactic flavor, while 
ours has a
semantic flavor. Moreover, our definition not only extends the 
concept of unfounded set to multi-valued logic, but also generalizes 
its definition to any given hypothesis $H$ (not just  the everywhere 
false hypothesis).

\begin{theorem}
Let $\cal P$ be a Datalog programs with negation.
\begin{enumerate}
\item  If $H_{\cal F}$ is the everywhere false hypothesis, then 
$sem_{\cal P}^{H_{\cal F}}$ coincides with the well-founded semantics 
of $\cal P$;
\item  If $H_{\cal U}$ is the everywhere underdefined hypothesis, 
then $sem_{\cal P}^{H_{\cal U}}$ coincides with the Kripke-Kleene 
semantics of $\cal P$.
\end{enumerate}
\end{theorem}
\preuve{
A Datalog programs with negation can be seen, in our  approach,
as a set of rules of the form

$$A \leftarrow (L_{1,1} \wedge ... \wedge L_{1,n}) \vee ... \vee
(L_{i,1}  \wedge ... \wedge L_{i,m}),$$
where the  $L_{p,q}$ are litterals.
\\\\
{\it First part.} 
Let $wfs(\cal P)$ be the well-founded semantics of $\cal P$.
We use in this part the definition of $wfs(\cal P)$
given in section  \ref{wfs}, considering all the
interpretations as functions from $\cal HB_P$ in $\{ \cal F,
U, T \}$.\\

Firstly, we show  that $wfs({\cal P}) =
T_R(wfs({\cal P})) \oplus s^{H_{\cal F}}_{\langle wfs({\cal P}), R
\rangle}$.\\\\
We know that $wfs({\cal P}) (A) = \cal T$ if and only if there is in $\cal 
P$ a 
rule 
$$A \leftarrow (L_{1,1} \wedge ... \wedge L_{1,n}) \vee ... \vee
(L_{i,1}  \wedge ... \wedge L_{i,m}), $$  such that there is 
$(L_{p,1} \wedge
... \wedge L_{p,q}) \in  \{ (L_{1,1} \wedge ... \wedge L_{1,n}) ; ... 
;
(L_{i,1}  \wedge ... \wedge L_{i,m}) \} $ such that $wfs({\cal P}) 
(L_{p,1}) =
... = wfs({\cal P}) (L_{p,q}) = \cal T$, i.e. if and only if 
$T_R(wfs({\cal P}))
(A) =  \cal T$.\\\\
We know that $wfs({\cal P}) (A) = \cal F$ if and only if  : 
\begin{itemize}
    \item  either $A$ is not the head of any rule of $R$, and then $s^{H_{\cal
F}}_{\langle wfs({\cal P}), R 
\rangle} (A) = \cal F$ ;
\item either there is in $\cal P$ a rule
$A \leftarrow (L_{1,1} \wedge ... \wedge L_{1,n}) \vee ... \vee 
(L_{i,1}  \wedge ... \wedge L_{i,m}), $  such that for all
$(L_{p,1} 
\wedge 
... \wedge L_{p,q}) \in  \{ (L_{1,1} \wedge ... \wedge L_{1,n}) ;
... ; 
(L_{i,1}  \wedge ... \wedge L_{i,m}) \} $  there is $L_{p,k} \in
\{L_{p,1}, 
... , L_{p,q} \}$ such that $wfs({\cal P} (L_{p,k}) = \cal F$,
i.e. if and only if   $T_R(wfs({\cal P}))
(A) =  \cal F$.
\end{itemize}
Consequently, $wfs({\cal P}) =
T_R(wfs({\cal P})) \oplus s^{H_{\cal F}}_{\langle wfs({\cal P}), R  
\rangle}$. It follows that $$sem_{\cal P}^{H_{\cal F}} \leq
wfs({\cal P}) .$$\\

We  prove now than $sem_{\cal P}^{H_{\cal F}} \leq
T^{\in}_{\cal P} (sem_{\cal P}^{H_{\cal F}}) \cup {\cal U}_{\cal P}
(sem_{\cal P}^{H_{\cal F}}) $.\\\\
We know that $sem_{\cal P}^{H_{\cal F}} (A) = \cal T$ if and only if 
there is in  $\cal P$ a rule
$$A \leftarrow (L_{1,1} \wedge ... \wedge L_{1,n}) \vee ... \vee 
(L_{i,1}  \wedge ... \wedge L_{i,m}), $$ such that $((L_{1,1} \wedge
... \wedge L_{1,n}) \vee ... \vee  
(L_{i,1}  \wedge ... \wedge L_{i,m})) \equiv {\cal T} ~w.r.t.~ sem_{\cal
P}^{H_{\cal F}}$ , i.e. such that there is  $(L_{p,1} \wedge
... \wedge L_{p,q}) \in  \{ (L_{1,1} \wedge ... \wedge L_{1,n}) ; ... 
;
(L_{i,1}  \wedge ... \wedge L_{i,m}) \} $ such that $ sem_{\cal
P}^{H_{\cal F}}(L_{p,1}) = 
... = sem_{\cal P}^{H_{\cal F}} (L_{p,q}) = \cal T$, i.e. if and only if
$T^{\in}_{\cal P} (sem_{\cal P}^{H_{\cal F}}) (A) = \cal
T$. \\\\
We know that $sem_{\cal P}^{H_{\cal F}} (A) = \cal F$ if and only if :
\begin{itemize}
    \item  either $A$ is not the head of any rule of $R$, 
    and then $ {\cal U}_{\cal P}
(sem_{\cal P}^{H_{\cal F}}) (A) = \cal F$; 
\item either there is in $\cal P$ a rule
$A \leftarrow (L_{1,1} \wedge ... \wedge L_{1,n}) \vee ... \vee 
(L_{i,1}  \wedge ... \wedge L_{i,m}), $  such that $((L_{1,1} \wedge
... \wedge L_{1,n}) \vee ... \vee  
(L_{i,1}  \wedge ... \wedge L_{i,m})) \equiv {\cal F} ~w.r.t.~sem_{\cal
P}^{H_{\cal F}}$, i.e. for all
$(L_{p,1} 
\wedge 
... \wedge L_{p,q}) \in  \{ (L_{1,1} \wedge ... \wedge L_{1,n}) ;
... ; 
(L_{i,1}  \wedge ... \wedge L_{i,m}) \} $  there is $L_{p,k} \in
\{L_{p,1}, 
... , L_{p,q} \}$ such that $sem_{\cal P}^{H_{\cal F}} (L_{p,k}) \equiv
\cal F$,  i.e. if and only if
  $ {\cal U}_{\cal P}
(sem_{\cal P}^{H_{\cal F}}) (A) = \cal F$.
\end{itemize}\vspace{.25cm}
Consequently, $sem_{\cal P}^{H_{\cal F}} \leq
T^{\in}_{\cal P} (sem_{\cal P}^{H_{\cal F}}) \cup {\cal U}_{\cal P}  
(sem_{\cal P}^{H_{\cal F}})$, thus $wfs({\cal P})   \leq  sem_{\cal 
P}^{H_{\cal F}}$.\\\\
{\it Second part.}  Let 
$I$ be an interpretation,  we show that $T_R (I) = {\Phi}_{\cal P} (I)$ 
where
${\Phi}_{\cal P}$ is the Kripke-Kleene operator. 

Then ${\Phi}_{\cal P} (I) (A) = \alpha \not = \cal U$ if and only if
there is in $\cal P$ a rule $A \leftarrow B$, where $B$
is defined by 
$ (L_{1,1} \wedge ... \wedge L_{1,n}) \vee ... \vee
(L_{i,1}  \wedge ... \wedge L_{i,m}) $, and if :
\begin{itemize}
    \item either there is
$(L_{j,1}  \wedge ... \wedge L_{j,k}) \subseteq B$ such that
$L_{j,1}  =  ...  =  L_{j,k} = \cal T$,
\item either for all $(L_{j,1}
\wedge ... \wedge L_{j,k}) \subseteq B$, there is $L_{j,l} \in \{
L_{j,1},  ..., L_{j,k} \}$ such that $L_{j,l} =\cal F$.
\end{itemize}
Thus
${\Phi}_{\cal P} (I) (A) = \alpha \not = \cal U$ if and only if
there is in $\cal P$ a rule $A \leftarrow B$ such that
$B \equiv_I \alpha$, i.e. if and only if $T_R(I) (A) = \alpha \not
= \cal U$.

As we consider the hypothesis $H_{\cal U}$, we will  infer 
 information only with $T_R$. $sem^{H_{\cal U}}_{\cal P}$ is the 
 least fixpoint of $T_R$ with respect to the knowledge ordering, thus
 $sem^{H_{\cal U}}_{\cal P}$ coincides with the
Kripke-Kleene semantics of $\cal P$ which is the least fixpoint of
${\Phi}_{\cal P}$ with respect to the knowledge ordering.

}

% The following proposition shows the relationship between the 
%different possible sound semantics of a program $\cal P$ for 
%different hypothesis $H$:

%\begin{proposition}
%Let $H_{\cal U}$ be the everywhere underdefined hypothesis and 
%$H_{\cal O}$  the everywhere overdefined hypothesis, then
%$s^{H_{{\cal U}}}_P\leq_k s^H_P\leq_k s^{H_{{\cal O}}}_P$ for all 
%hypothesis $H$.
%\end{proposition}

\section{Concluding remarks}

We have defined a formal framework for information integration based 
on hypothesis testing. A basic concept of this framework is the 
support provided by a program ${\cal P} = \langle F, R \rangle$ to a 
hypothesis $H$. The support of $H$ is the maximal part of $H$ that 
does not contradict the facts of $F$ or the facts derived from $F$ 
using the rules of $R$.

We have then used the concept of support to define hypothesis-based 
semantics for the class of Fitting programs, and we have given an 
algorithm for computing these semantics.

Finally, we have shown that our  semantics extends the well-founded 
semantics and the Kripke-Kleene semantics to multi-valued 
logics with bilattice structure, and also generalizes them
in the following sense: if we 
restrict our attention to three-valued logics then for $H_{\cal F}$  
the everywhere $false$ interpretation our semantics reduces to the 
well-founded semantics, and for $H_{\cal U}$  the everywhere 
$underdefined$ interpretation  our semantics reduces to the 
Kripke-Kleene  semantics.

We believe that hypothesis-based semantics can be useful not only in 
the context of information integration but also in the context of 
explanation-based systems. Indeed,  assume that a given hypothesis 
$H$ turns out to be a part of the $H$-semantics of a program $\cal 
P$. Then  $\cal P$ can be seen as an ``explanation'' of the 
hypothesis $H$. We are currently investigating several aspects of 
this explanation oriented viewpoint.

\end{document}